\documentclass{iau}
\usepackage{graphicx}

\usepackage{txfonts}
\usepackage{color}
\usepackage{colortab}
\usepackage{pstricks}
\usepackage{enumerate}
\usepackage[normalem]{ulem}

\title[Multi-resonance orbital model of HF~QPOs] 
{Multi-resonance orbital model of HF~QPOs}

\author[Zden\v{e}k Stuchl\'{\i}k, Andrea Kotrlov\'a \& Gabriel T\"{o}r\"{o}k]   
{Zden\v{e}k Stuchl\'{\i}k, Andrea Kotrlov\'a \and Gabriel T\"{o}r\"{o}k}

\affiliation{Institute of Physics, Faculty of Philosophy and Science, Silesian University in Opava,\\
Bezru\v{c}ovo n\'{a}m. 13, CZ-74601 Opava, Czech Republic\\ email: {\tt zdenek.stuchlik@fpf.slu.cz}, {\tt andrea.kotrlova@fpf.slu.cz} \\[\affilskip]
}

\pubyear{2012}
\volume{290}  
\jname{Feeding compact objects: Accretion on all scales}
\editors{C.M. Zhang, T. Belloni, M. M\'endez \& S.N. Zhang, eds.}

\begin{document}

\maketitle

\begin{abstract}
Using known frequencies of the~twin peak high-frequency quasiperiodic oscillations (HF~QPOs) and known mass $M$ of the~central black hole, the~black-hole dimensionless spin $a$ can be determined assuming a~concrete version of the~resonance model. However, large range of observationally limited values of the~black hole mass implies a~low precision of the~spin estimates. We discuss the~possibility of higher precision of the~black hole spin $a$ measurements in the~framework of multi-resonance model inspired by observations of more than two HF~QPOs in some black hole sources. We determine the~spin and mass dependence of the~twin peak frequencies with a~general rational ratio $n:m$ assuming a~non-linear resonance of oscillations with the~epicyclic and Keplerian frequencies or their combinations. In the~multi-resonant model, the~twin peak resonances are combined properly to give the~observed frequency set. We focus on the~special case of duplex frequencies, when the~top, bottom, or mixed frequency is common at two different radii where the~resonances occur giving triple frequency sets.
\end{abstract}

\firstsection 
\section{Multi-resonance models with Keplerian and epicyclic oscillations}

The~standard orbital resonance model assumes non-linear resonance between oscillation modes of an~accretion disc orbiting a~central object, here considered to be a~rotating Kerr black hole. The~frequency of the~oscillations is related to the~Keplerian frequency  $\nu_\mathrm{K}$ (orbital frequency of tori), or the~radial $\nu_{r}$ and vertical $\nu_{\theta}$  epicyclic frequencies of the~circular test particle motion.


\subsection{More resonances sharing one specific radius}

This special case allows existence of so called strong resonant phenomena when two (or more) versions of resonance could occur at the~same radius allowing cooperative effects between the~resonances~\cite{Stu-Kot-Tor:2008:ACTA:BHadmStrResPhen}. Of course, such a situation is allowed for black holes with a specific spin only. Of special interest seems to be the~case of the~``magic'' spin $a=0.983$, when the~Keplerian and epicyclic frequencies are in the~ratio $\nu_{\mathrm{K}}\!:\!\nu_{\theta}\!:\!\nu_{r} = 3\!:\!2\!:\!1$ at the~common radius $x\equiv r/M = 2.395$, see Fig.\,\ref{fig1}.

\subsection{Resonances occurring at two specific radii -- triple frequency sets}

\begin{figure}[t]
\begin{center}
 \includegraphics[width=2.6in]{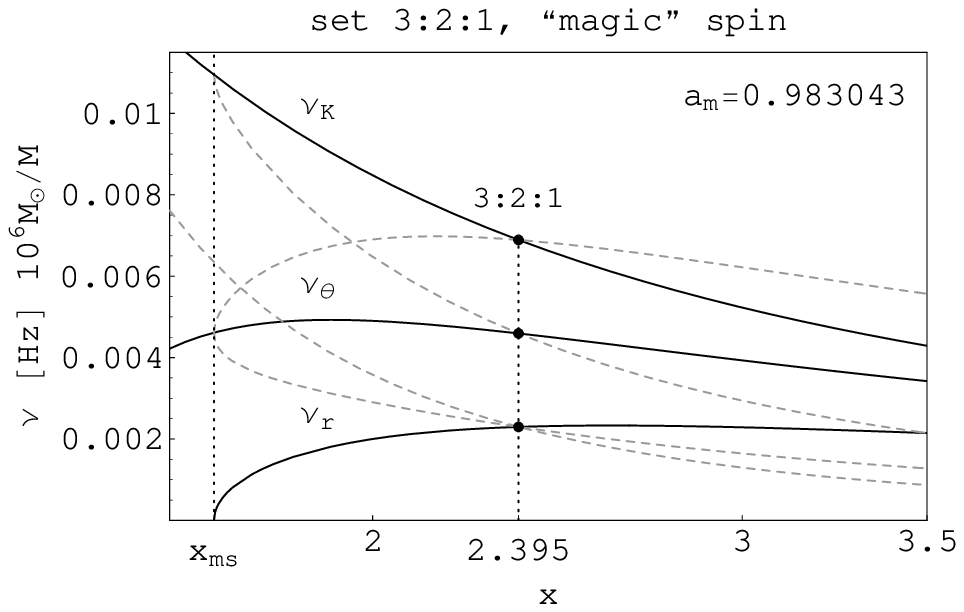}\hfill
 \includegraphics[width=2.6in]{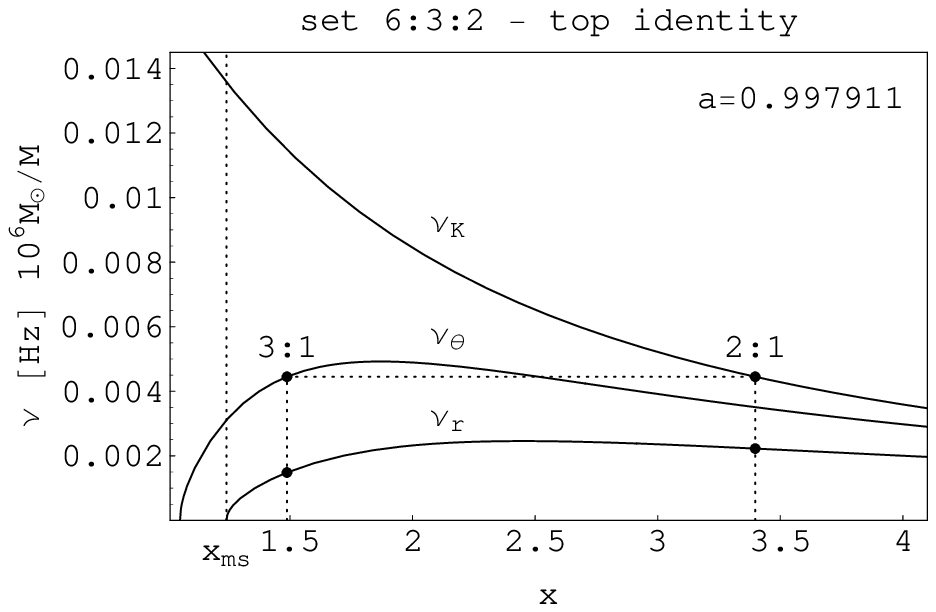}
 \caption{\textit{Left:} The special case of a ``magic'' spin,
  when the strongest resonances could occur at the same radius. For
  completeness we present the relevant simple combinational frequencies
  $\nu_{\theta}-\nu_{r}$, $\nu_{\theta}+\nu_{r}$,
  $\nu_{\mathrm{K}}-\nu_{\theta}$, $\nu_{\mathrm{K}}-\nu_{r}$ (grey dashed
  lines). \textit{Right:} The case of the duplex frequencies when the top frequency is common at two different radii where the resonances occur giving triple frequency ratio set.}
   \label{fig1}
\end{center}
\end{figure}

 In general, we can expect the
  oscillations to be excited at two different radii of the accretion
  disc and to enter the resonance in the framework of
  different versions of the resonance model (i.e., four frequency set is observed generally). 
  In special cases, for some specific values of the~black hole spin, two twin peak QPOs observed at the~radii $x_{n:m}$ and $x_{n':m'}$ have the~top (see Fig.\,\ref{fig1}), bottom or mixed (the~bottom at the~inner radius and the~top in the~outer radius, or vice versa) frequencies identical~\cite[(Stuchl{\'{\i}}k {et~al.} 2012)]{Stu-Kot-Tor:2011:AA}. Such situations can be characterized by sets of only three frequencies (upper $\nu_{\mathrm{U}}$, middle $\nu_{\mathrm{M}}$ and lower $\nu_{\mathrm{L}}$) with ratio
$\nu_{\mathrm{U}}:\nu_{\mathrm{M}}:\nu_{\mathrm{L}} = s:t:u$.

Let us consider a~simple situation with the~``top identity'' of the~upper frequencies in two direct resonances between the~radial $\nu_{r}$ and vertical $\nu_{\theta}$ epicyclic oscillations at two different radii $x_{p},\,x_{p'}$ with $p^{1/2}=m\!:\!n$, $p'^{1/2}=m'\!:\!n'$. The~condition
$\nu_{\theta}(a,x_p)=\nu_{\theta}(a,x_{p'})$ is then transformed to the~relation
\begin{equation}
\alpha_{\theta}^{1/2}(a,x_{p}) \left(x_{p}^{3/2}+a\right)^{-1} =
\alpha_{\theta}^{1/2}(a,x_{p'}) \left(x_{p'}^{3/2}+a\right)^{-1}
\end{equation}
which uniquely determines the~black hole spin $a$, since the~radii $x_{p}$ and $x_{p'}$ are related to the~spin $a$ by the~resonance conditions
   \begin{equation}\label{aD1}
          a=a^{\theta/{r}}(x,p)\equiv\frac{\sqrt{x}}{3(p+1)}\left\{2(p+2)-\sqrt{(1-p)\left[3x(p+1)-2(2p+1)\right]}\right\}
    \end{equation}
for $a^{\theta/{r}}(x,p)$ and $a^{\theta/{r}}(x,p')$, respectively. When two different resonances are combined, we proceed in the~same manner~\cite[(Stuchl{\'{\i}}k {et~al.} 2012)]{Stu-Kot-Tor:2011:AA}. The~black hole spin $a$ is given by the~types of the~two resonances and the~ratios $p$, $p'$, quite independently of the~black hole mass $M$. A detailed table guide across all the possible triple frequency sets and
related values of the black hole spin $a$ (limited by $n
\leq 4$) is presented in~\cite{Stu-Kot-Tor:2011:AA}.

\section{Conclusions}

The~multi-resonance model of HF~QPOs can be considered as a promising approach to understand the~observational data from black holes sources. The special triple frequency set method determines the
black hole spin precisely, but not uniquely, as in general the same frequency set could occur for different values of the spin within different versions of the resonance model. In such situations the black hole spin estimates coming from the spectra fitting and the line profile model could be relevant in determining the proper versions of the resonant model. When the black hole spin is found, its mass can be determined from the magnitude of the observed frequencies. The efficiency of the black hole spin determination by using the
triple frequency set ratios grows strongly with growing precision
of the frequency measurements. The~prepared new space X-ray mission LOFT proposes sensibility of the~observational instruments high enough to reach data that could be precise enough to make application of the~triple frequency set method realistic.
\\\\
\textbf{Acknowledgements.} The authors acknowledge the research grant GA\v{C}R~202/09/0772 and the project CZ.1.07/2.3.00/20.0071 ``Synergy'' supporting international collaboration of the Institute of Physics at SU Opava.

%

\end{document}